\documentclass[%
 reprint,
 amsmath,amssymb,
 aps,
prb,
]{revtex4-1}
\setcitestyle{numbers,square}

\usepackage{braket}
\usepackage{graphicx}
\usepackage{dcolumn}
\usepackage{bm}
\usepackage[normalem]{ulem} 
\usepackage{color}

\usepackage{hyperref}
\hypersetup{
    colorlinks,%
    citecolor=blue,%
    linkcolor=blue,%
    urlcolor=blue
}

\newcommand{\cinco}{Si$_2$O$_5$}

\begin{document}

\title{Engineering metal-$sp_{xy}$ Dirac bands on the oxidized SiC surface}

\author{F. Crasto de Lima}
\email{felipe.lima@ufu.br}
\author{R. H. Miwa}
\email{hiroki@ufu.br}
\affiliation{Instituto de F{\'i}sica, Universidade Federal de Uberl{\^a}ndia, \\
       C.P. 593, 38400-902, Uberl{\^a}ndia, MG,  Brazil}%
\date{\today}

\begin{abstract} 

The ability to construct 2D systems, beyond materials natural formation,
enriches the search and control capability of new phenomena. For 
instance, the synthesis of topological lattices of vacancies on metal surfaces 
through scanning tunneling microscopy. In the present study we demonstrate 
that metal atoms encaged in silicate adlayer on silicon carbide is an 
interesting platform for lattices design, providing a ground to experimentally 
construct tight-binding models on an insulating substrate. Based on the density 
functional theory, we have characterized the energetic and the electronic 
properties of 2D metal lattices embedded in the silica adlayer. We show that the 
characteristic band structures of those lattices are ruled by surface states 
induced by the metal-$s$ orbitals coupled by the host-$p_{xy}$ states; 
giving rise to $sp_{xy}$ Dirac bands neatly lying within the energy gap of the 
semiconductor substrate.

\end{abstract}

\maketitle

In recent years two-dimensional (2D) materials have emerged with prominent phenomena and applications. For instance, graphene, the first observed 2D material, presents relativistic quasiparticles described by the massless Dirac equation \cite{NATUREGeim2007}; meanwhile many other materials have been theoretically predicted \cite{NATUREMounet2018, 2DMHaastrup2018} and experimentally synthesized \cite{AMGeng2018}. Within these, new quasiparticles \cite{PRBLan2011, JPCLWang2018, PRLWang2019, PRBcrasto2020}, and topological/semimetal phases \cite{NATURELu2014, PRBZhu2018, NATUREGao2018} have attracted great interest in fundamental physics. Focusing in technological applications, quantum Hall effects (spin \cite{PRLKane2005, SCIENCEHatsuda2018}, anomalous \cite{PRBOstrovsky2008, PCCPdeLima2018} and valley \cite{SCIENCEKomatsu2018}) and thermoelectric properties, to cite a few, are  explored for devices engineering based on 2D systems \cite{NATURESnyder2008, PRBAng2017, CMNaghavi2018, PRMDong2019, JCPCrasto2019}.

The ability to construct lattices on demand promotes the exploration/enhancement of the materials properties. Currently, we are facing a striking synergy between theoretical exploration and experimental lattices designs. For instance, artificial graphene has been constructed in quantum dots systems \cite{PRLrasanen2012} and in 2D electron gas \cite{PRBgibertini2009} allowing the control of the Dirac quasiparticles and topological phases \cite{NATUREpolini2013}. Beyond these systems, within the organic chemistry,  covalent organic frameworks and metal organic frameworks have been successfully synthesized by combining different molecules/metal centers \cite{Colson2013, CCRodriguez2016, NATUREBaumann2019}. 

Further control over lattice formation has been exploited through scanning tunneling microscopy (STM) techniques ``printing'' atom-by-atom on solid surfaces, where the STM tip  brings precise control over the lattice formation \cite{SCIENCECrommie1993, RSIcelotta2014, NATUREPavlicek2017, PRXSlot2019}. For instance, topological states have been engineered on atomic square lattice in chlorine covered Cu(100) surface by vacancy formation \cite{NATUREDrost2017}. Changing the substrate surface to Cu(111) lieb lattice \cite{NATURESlot2017}, graphene-like \cite{NATUREGomes2012} and quasicristals \cite{NATURECollins2017} have been imprinted in a carbon monoxide cover layer, while fractal geometric lattices have been achieved in Co covered Cu(111) \cite{NATUREKempkes2019}. Those studies took advantage of the current state of the art on the control over the atomic and molecular adsorption/desorption processes on metal surfaces. However, it is worth pointing out that,  eventually  (i) the presence of the (host) metallic states may blur the characteristic band structure of a given designed lattice, and/or (ii) such a surface supported 2D lattices not being stable/robust against temperature.

Silicon carbide (SiC) has been considered a promising semiconductor material for 
applications in electronic and biological devices. It presents outstanding high 
temperature, voltage and power  stability \cite{CRSSMSVolker2008}. Additionally, 
many of silica ordered phases \cite{JPCCwang2015} emerged as a crystalline 2D 
insulator where metals ions can be encapsulated in its pores 
\cite{JPCCbuchner2014, PSSchristin2017}. Upon surface oxidation, highly-ordered 
2D silica has been experimentally shown to form on the SiC surface 
\cite{APLbernhardt1999, PRBSchurmann2006, PRBHoshino2004, APLTochihara2014}, 
while in a recent study metal bounded to the SiC surface, within the silica 
adlayer has been explored \cite{CARBONPadilha2019}. This system presents a 
semiconducting surface, while providing additional protection against external 
perturbation/interfaces as the metal are caged by the 2D silicate (Si$_2$O$_5$) 
structure; and thus a quite interesting system overcome the issues (i) and (ii) 
above.

\begin{figure}[h!]
\includegraphics[width=\columnwidth]{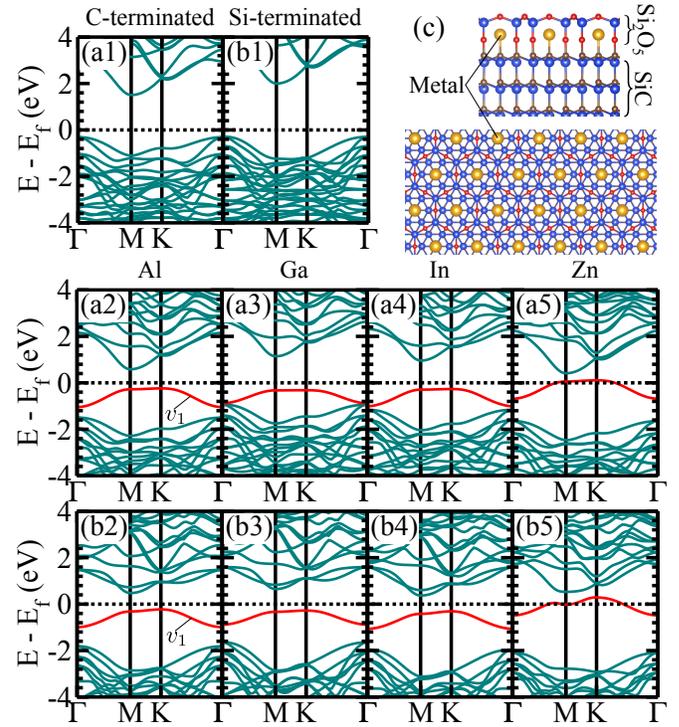}
\caption{\label{band-tri} Band structure for the pristine \cinco/SiC (a1)-(b1), and with Al (a2)-(b2), Ga (a3)-(b3), In (a4)-(b4), and Zn (a5)-(b5) intercalated metals in all vacant sites, i.e. triangular lattice shown in (c). The C- and Si-termination of SiC are shown in panels (a) and (b), respectively. The red lines indicate the metals states in the middle of the (M)\cinco/SiC valence and conduction bands, which are shown in blue.}
\end{figure}

In this paper we explore the design of 2D lattices on the metal adsorbed 
oxidized silicon carbide surfaces. The   metal (M) adatoms, with M=Al, Ga, In, 
and Zn,  are patterned by the   highly-ordered 2D silicate adlayer on the SiC 
surface, \cinco/SiC \cite{APLbernhardt1999, PRBSchurmann2006, PRBHoshino2004, 
APLTochihara2014}, giving rise to  triangular metal lattices  upon the formation 
of metal--surface  chemical bonds embedded in \cinco, (M)\cinco/SiC, 
Fig.~\ref{band-tri}. We show that the 2D lattices can be constructed by the 
creation of metal vacancies on the surface. In particular, we have examined four 
2D lattices, {\it viz.}: honeycomb, kagome and two lattices with higher 
pseudospin Dirac quasiparticles, however, our findings are not limited to those 
ones. Our electronic structure results reveal that the characteristic band 
structures of the 2D lattices are dictated by the metal-$s$ states (dominant 
contribution) hybridized with  the host-$p_{x,y}$ orbitals, giving rise to 
$sp_{x,y}$ Dirac bands neatly lying in the band gap of a semiconductor surface.


\paragraph*{{\bf Results and Discussions.} Silicate adlayer metallization.} We 
have considered oxidized SiC surfaces forming a 
($\sqrt{3}\times\sqrt{3}$)$R30^\circ$ reconstructed {\cinco} silicate adlayer on 
the C [Si] terminated surface, \cinco/SiC(000$\bar 1$) [\cinco/SiC(0001)]. The 
silicate/SiC interface is characterized by a hexagonal structure of O--C [O--Si] 
chemical bonds with the topmost atoms of the (000$\bar1$) [(0001)] surface, 
centered by a single C [Si] dangling bond per surface unit cell. The unpaired 
electron of the dangling bond  gives rise to localized spin-polarized surface 
states (SSs) within the energy gap of SiC \cite{PRBlu2000}. As shown in 
Fig.~\ref{band-tri}(a1) and (b1), such SSs are washed out from the energy gap by 
the hydrogen passivation of those dangling bonds. We found a gap of $1.81$ and 
$2.29$\,eV for C and Si terminations, respectively; consistent with previous 
calculations \cite{PRBlu2000, CARBONPadilha2019}.

Instead of hydrogen, the passivation of the surface dangling bonds by post-transition metals, Al, Ga, In and Zn [Fig.~\ref{band-tri}(c)] will bring back the SSs; but now ruled by spherically symmetric $s$ orbitals of the metal adatoms hybridized with the host \cinco/SiC surface. For instance, the passivation of the C [Si] dangling bond by an Al adatom, (Al)\cinco/SiC(000$\bar 1$) [(Al)\cinco/SiC(0001)], the Al-$3p^1$ electron will passivate the dangling bond. Meanwhile the Al-$3s^2$ orbitals will give rise to a fully occupied SS lying within the band gap of SiC; indicated as $v1$ in Figs.~\ref{band-tri}(a2) and \ref{band-tri}(b2) for (Al)\cinco/SiC(000$\bar 1$) and (Al)\cinco/SiC(0001), respectively. It is worth noting that $v1$ presents an energy dispersion of about 1\,eV for wave vectors parallel to the surface, indicating an electronic interaction between the Al adatoms mediated by the \cinco/SiC host orbitals. Indeed, through the calculation of the  projected electronic band structure (not shown) we find that the $3s$-orbital of the Al adatom hybridizes with the $p_z$ orbital of the surface (Si or C) dangling bond, and planar $p_{x,y}$ orbitals of the silicate layer. The latter rules for the electronic interaction between the nearest-neighbor (NN) Al adatoms. The same behavior has been observed for the same valence Ga, and In metals, 
Figs.~\ref{band-tri}(a3)-(b3) and ~\ref{band-tri}(a4)-(b4) respectively. Moreover, for Zn atoms, which present one valence electron less than Al, a partially filled 4$s$ state is predicted on the Fermi energy, as shown in Fig.~\ref{band-tri}(a5)-(b5). 

\begin{figure}[h!]
\includegraphics[width=\columnwidth]{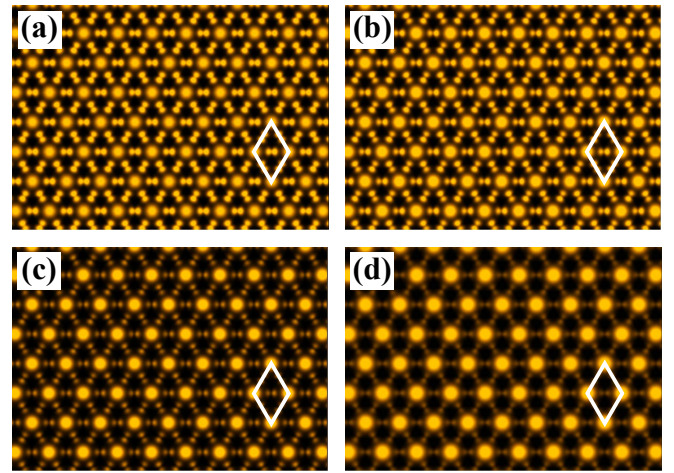}
\caption{\label{STM-triangular} Constant height STM simulation, within an energy interval of 1 eV below the Fermi level, for fully metallized (triangular lattice) (M)\cinco/SiC(000$\bar 1$) for M = Al (a), Ga (b), In (c) and Zn (d).} 
\end{figure}

Further electronic structure characterization was performed by a set of STM 
simulations of the occupied states, shown in Fig.~\ref{STM-triangular}. Although 
the metal atoms lie below the topmost Si--O hexagonal lattice of the silicate 
layer, the simulated images are characterized by bright spots lying on the metal 
adatoms forming a triangular lattice connected by less intense bright lines. We 
are picturing the metal induced ($s$-orbital) SSs, hybridized with the surface  
planar $p_{x,y}$ orbitals, giving rise to ``electronic bonds''  connecting the 
NN metal atoms.

\begin{table}[h!]
\begin{ruledtabular}
\caption{\label{tab1} Vacancy formation energy per exchanged M$\leftrightarrows$H atoms $\Delta E_{\eta}$ (eV) for the honeycomb, kagome, triangular-rectangular (tri-rect) and snub-hexagonal (snub-hex) lattices.}
\begin{tabular}{ccccc}
 & \multicolumn{4}{c}{M/C-SiC(O)} \\
 \cline{2-5}
Configuration    &  Al  &  Ga  &  In  & Zn \\  
\hline
honeycomb ($1/3$) & 1.17 & 1.33 & 1.86 & 2.91 \\
kagome    ($1/4$) & 1.43 & 1.61 & 2.22 & 3.41 \\
tri-rect  ($1/4$) & 1.43 & 1.61 & 2.22 & 3.41 \\
snub-hex  ($1/7$) & 1.78 & 1.98 & 2.69 & 4.05 \\
\hline
 & \multicolumn{4}{c}{M/Si-SiC(O)} \\
 \cline{2-5}
Configuration    &  Al  &  Ga  &  In  &  Zn  \\
\hline
honeycomb ($1/3$) & 1.86 & 1.87 & 2.38 & 2.99 \\
kagome    ($1/4$) & 2.07 & 2.22 & 2.81 & 3.50 \\
tri-rect  ($1/4$) & 2.07 & 2.22 & 2.81 & 3.50 \\
snub-hex  ($1/7$) & 2.66 & 2.68 & 3.36 & 4.14 \\
\end{tabular}
\end{ruledtabular}
\end{table}


\paragraph*{Design of 2D lattices.} Given such an electronic scenario, based on the SSs induced by the metal $s$-orbital in the band gap of SiC, (M)\cinco/SiC is an interesting and realistic semiconductor platform to design 2D lattices throughout the creation of metal vacancies on the surface \cite{NATUREDrost2017,yan2019engineered}. Indeed, in a recent study, we explored the electronic properties of Archimedean lattice models based in a 2D $s$-orbital tight-binding model \cite{PCCPdeLima2018}. Here, we explore the electronic properties of triangular based lattices where the metal vacancy sites are saturated by hydrogen atoms, in order to preserve the semiconducting character of the \cinco/SiC host surface. 

Since the proposed lattice design is based on an atomic exchange between the metal atoms, embedded in (M)\cinco/SiC, by  hydrogen atoms passivating the remnant surface dangling bonds, the energetic stability of the final system was inferred through the following total energy comparison ($\Delta E_\eta$),
\begin{equation}
\Delta E_\eta = E_\eta - E_{\eta = 0} + n\times(E_{\rm M} - E_{\rm H}),
\end{equation}
where $E_\eta$ is the total energy of  the (M)\cinco/SiC surface with a given concentration ($\eta$) of metal atoms exchanged by hydrogen atoms (M$\leftrightarrows$H), $n$ is the number M$\leftrightarrows$H exchanged atoms, $\eta=n/n_{\rm tot}$ ($n_{\rm tot}$ is the total number of sites); while $E_{\eta=0}$ is the total energy of the fully metallized surface, and $E_{\rm M}$ ($E_{\rm H}$) is the total energies of an isolated metal (hydrogen) atom. Our results of $\Delta E_\eta$, summarized in Table\,\ref{tab1}, indicate that the M$\leftrightarrows$H atomic exchange processes are exothermic  ($\Delta E_\eta > 0$), and thus supporting the feasibility of engineering 2D lattices on the (M)\cinco/SiC semiconductor surfaces. Indeed, very recently, 2D atomic lattices based on ``vacancy design'' have been successfully synthesized on metal surface, clorine-monolayer/Cu(100) \cite{NATUREDrost2017}.

\begin{figure}[h!]
\includegraphics[width=\columnwidth]{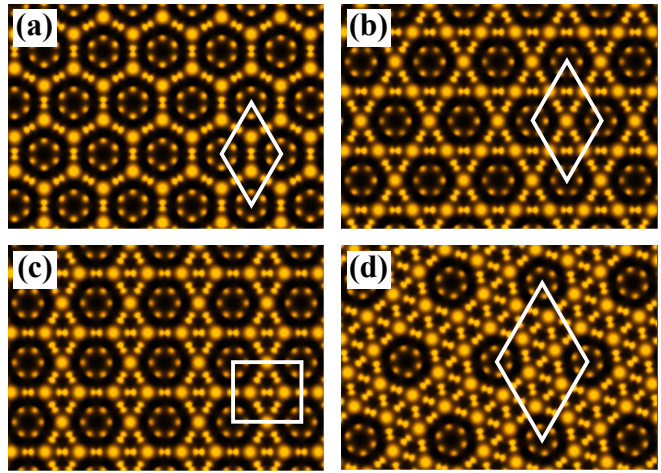}
\caption{\label{stm} Constant height STM, within an energy interval of 1 eV below the Fermi level, for (Al)\cinco/SiC(000$\bar 1$) in the  honeycomb (a) kagome (b), triangular-rectangular (c) and snub-hexagonal (d) lattices.}
\end{figure}

In order to illustrate the lattice formation, in Fig.~\ref{stm} we present the simulated STM images of the occupied states, within an energy interval of 1\,eV below the Fermi level of the (Al)\cinco/SiC(000$\bar 1$) surface. Here, we have considered the following triangular based lattices, honeycomb ($\eta$=1/3), kagome ($\eta$=1/4), triangular-rectangular ($\eta$=1/4), and snub-hexagonal ($\eta$=1/7), as depicted  in Figs.~\ref{stm}(a)--\ref{stm}(d). It is worth noting that other 2D lattices can be designed on the (M)\cinco/SiC surface, for instance, the ones predicted/synthesized based on boron atoms \cite{NLPenev2012, ACIEZhang2015, NLCrasto2019}.

The STM pictures are characterized by dark spots on the Al$\leftrightarrows$H exchanged sites, since  the electronic states of the C--H bonds are resonant within the valence and conduction  bands of SiC. In contrast, the Al atoms are identified by the SSs composed Al-$3s$ orbitals hybridized with the surface host orbitals. In this case, we find bright spots on the Al atoms, connected by ``electronic bonds'', forming an artificial Al lattice on the \cinco/SiC(000$\bar 1$) surface. Similar STM results were obtained for the other (M)\cinco/SiC systems.

We can estimate the inter-metal coupling due the ``electronic bonds'' by comparing the hopping strength ($t$) with the band width from a NN tight-binding model \cite{njpBENA2009}. The coupling between the metals are insensitive to the surface termination, and the difference in $t$ by less then $5$\,meV for same lattice and different terminations. We find $t = 101,\, 70,\, 90,$ and $97$\,meV, for M = Al, Ga, In and Zn, respectively. These values are close to the experimentally achieved in the interacting lattice of chlorine layer on Cu(100) \cite{NATUREDrost2017}. Such an interaction plays an important role on the electronic band structure of the 2D lattices on (M)\cinco/SiC.

\begin{figure}[h!]
\includegraphics[width=\columnwidth]{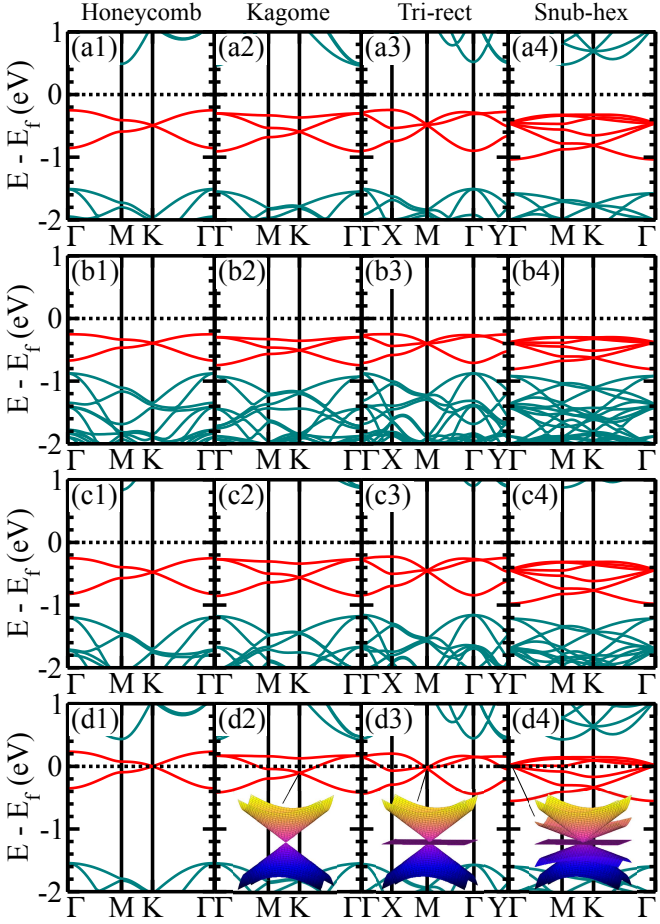}
\caption{\label{band-c} Lattices band structure in (M)\cinco/SiC(000$\bar 1$) for (a) M=Al, (b) M=Ga, (c) M=In and (d) M=Zn. (a1)-(d1) Honeycomb lattice, (a2)-(d2) kagome lattice, (a3)-(d3) tri-rectangular lattice and (a4)-(d4) snub-hexagonal lattice. Red lines indicate the metal states in the middle of SiC(O) valence and conduction bands, which are shown in blue.}
\end{figure}

\begin{figure}[t]
\includegraphics[width=\columnwidth]{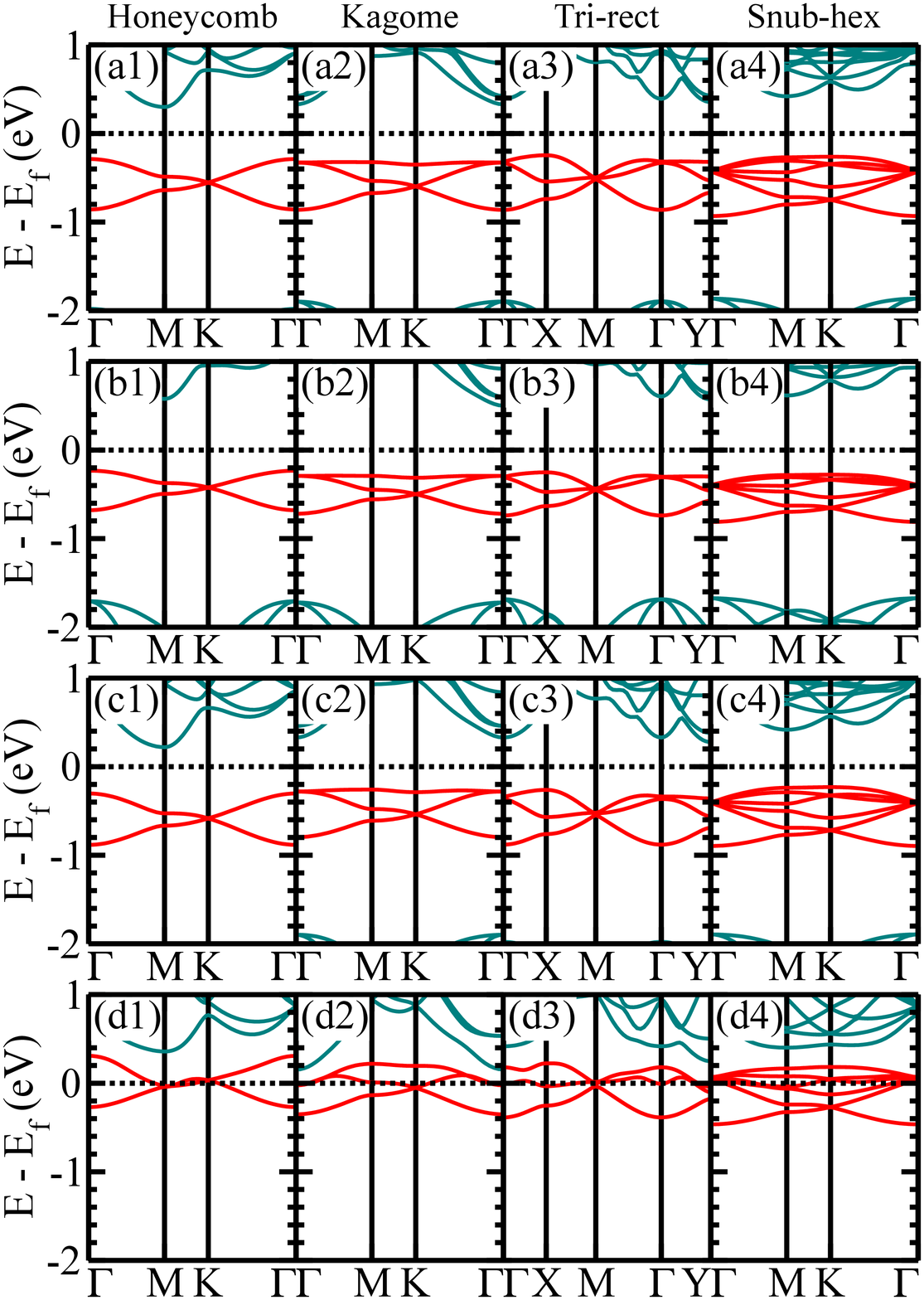}
\caption{\label{band-si} Lattices band structure in (M)\cinco/SiC(0001) for (a) M = Al, (b) M = Ga, (c) M = In and (d) M = Zn. (a1)-(d1) Honeycomb lattice, (a2)-(d2) kagome lattice, (a3)-(d3) tri-rectangular lattice and (a4)-(d4) snub-hexagonal lattice. Red lines indicate the metal states in the middle of SiC(O) valence and conduction bands, which are shown in blue.}
\end{figure}

\paragraph*{Lattice's electronic structure.} The electronic band structures of the (M)\cinco/SiC surfaces, upon the M$\leftrightarrows$H atomic exchanges, can be considered as the fingerprints of the (proposed) artificial metal lattices. In panel (a1) of Figs.~\ref{band-c} and \ref{band-si} we present the electronic band structure of (Al)\cinco/SiC(000$\bar 1$) and /SiC(0001) surfaces, where we have an artificial graphene-like structure of Al atoms. We find linear energy dispersions at the K-point, and the projection of those energy bands reveals that those Dirac bands are mostly composed by Al-$3s$ orbitals hybridized with (i) the planar O-$2p_{x,y}$ orbitals of silicate layers, and (ii) the $p_z$ orbitals of the topmost C/Si atoms of the SiC(000$\bar 1$)/(0001) surface bonded to Al adatoms. The energy dispersion of those Dirac bands are  mostly dictated by the overlap of Al-$3s$ and O-$2p_{x,y}$ orbitals [(i)]. Kagome energy bands, {\it i.e.} graphenelike energy bands forming a Dirac cone at the K-point,  degenerated with a nearly flat band at the center of the Brillouin zone, are nicely reproduced for a artificial kagome lattice of Al atoms, panel (a2) of Figs.~\ref{band-c} and \ref{band-si}. Additionally, by constructing different lattices, we could achieve the pseudospin-1 and -2 Dirac quasiparticles \cite{PCCPCrasto2019, PRBcrasto2020} at the M point and $\Gamma$ point of triangular-rectangular and snub-hexagonal lattices, respectively shown in panels (a3)-(c3) and (a4)-(c4) of Fig.~\ref{band-c} and Fig.~\ref{band-si}. Different from the spin-$1/2$ Dirac equation, leading to the Dirac cone [see inset of Fig.~\ref{band-c}(d2)], these quasiparticles generalize the Pauli matrix in the Dirac equation by the angular momentum matrix of higher spin \cite{PRBurban2011, RESEARCHfang2019}. We can characterize the pseudospin-1 Dirac quasiparticle by a Dirac cone degenerated with a flat band [see inset of Fig.~\ref{band-c}(d3)], while for the pseudospin-2 case two Dirac cones with velocities related by a factor of 2 are degenerated with a flat band [see inset of Fig.~\ref{band-c}(d4)].

Since Ga and In atoms exhibit the same $s^2p^1$ valence configuration, 
(Ga)\cinco/ and (In)\cinco/SiC  exhibit nearly the same characteristic band 
structures as depicted in panels (a2)-(d2) and (a3)-(d4) of Figs.~\ref{band-c} 
and \ref{band-si}. In both (M)\cinco/SiC systems,  the characteristic bands are 
localized below the Fermi energy. Whereas, in (Zn)\cinco/SiC the Zn-$4s$ orbital 
gives rise to metallic bands, panels (d1)-(d4) in Fig.~\ref{band-c} and 
Fig.~\ref{band-si}. In particular, for the Si terminated (Zn)\cinco/SiC(0001) 
surface [Figs.~\ref{band-si}(d1)-(d4)], the Zn-$4s$ states are 
nearly resonant with the  the  edge of the SiC bulk conduction band. In 
this case, the energy dispersions of the characteristic bands  become distorted, 
(i) the honeycomb lattice present additional crossing along M-K points besides 
the Dirac states at K point [Fig.~\ref{band-si}(d1)], (ii) the kagome flat band 
bend upward [Fig.~\ref{band-si}(d2)], and (iii) degeneracy breaking at $\Gamma$ 
point of triangular-rectangular and snub-hexagonal lattice are observed 
[Fig.~\ref{band-si}(d3) and (d4)]. It is worth pointing out that we have 
performed calculations with spin-orbit coupling (SOC) in order to clarify its 
effect in the band structures. Here we found a small SOC effect, where due to 
the surface inversion symmetry breaking, a Rashba spliting from $0.01\%$ to 
$6\%$ of the band width been found with the highest value for 
(In)\cinco/SiC(0001). Such scenario does not alter significantly the electronic 
structure of the systems, which preserve its characteristic band dispersion.

 
 In summary, we have explored the design of 2D lattices based on 
incorporation/removal of metal (M) adatoms patterned by ordered silicate adlayer 
on the SiC surface, (M)\cinco/SiC. Here, the metal adatoms are caged (protected) 
within the silicate adlayer, where we demonstrate that the metals can still be 
observed in STM experiments. We shown that Al, Ga, In and Zn metal atoms 
present their valence $s$ states within the substrate energy gap, allowing an 
exploration of lattices' electronic structure. As a proof of principle, we 
characterize the arising (i) relativistic Dirac states in a honeycomb lattice, 
(ii) flat bands in a kagome lattice, and (iii) pseudospin-1 and 2 in a 
triangular-rectangular and snub-hexagonal lattices, respectively; however, our 
findings can be extended to other 2D lattices on the \cinco/SiC surface.

\vspace*{0.2cm}

\paragraph*{\bf Computational approach.} We have performed {\it first-principle} calculations, within density functional theory, as implemented in the plane wave based VASP code \cite{CMSKresse1996}. The exchange and correlation was described by Perdew, Burke, and Ernzerhof (PBE) functional \cite{PRLPerdew1996}, with the plane wave basis with 400 eV cutoff energy. We have considered the SiC(0001) and SiC(000$\bar{1}$) slabs with six atomic layers, with one side bonded to the silica Si$_2$O$_5$ and the other hydrogen passivated. The electron-ion interaction was taken into account throughout the projector augmented wave method (PAW) \cite{PRBBlochl1994}. All atoms positions have been relaxed until all forces were less then $5$\,meV/{\AA}, within a k-mesh of $7/N\times 7/N\times 1$ special points \cite{PRBmonkhorst1976}, where $N$ is the supercell multiplicity (for the unity cell $N=1$). For the band structure calculations a increased k-mesh of $11/N\times 11/N \times 1$ have been considered.

\section*{ACKNOWLEDGMENTS}

The authors acknowledge financial support from the Brazilian agencies CNPq, CAPES, and FAPEMIG, and the CENAPAD-SP and Laborat{\'o}rio Nacional de Computa{\c{c}}{\~a}o Cient{\'i}fica (LNCC-SCAFMat2) for computer time. 

\bibliography{bib}

\end{document}